\documentclass[12pt]{iopart}

\usepackage{latexsym}
\usepackage{graphicx}
\usepackage{color}

\begin{document}

\title{Electron glass signatures up to room temperature in disordered insulators}

\author{J. Delahaye and T. Grenet}

\address{Univ. Grenoble Alpes, CNRS, Institut Néel, 38000 Grenoble, France}
\ead{julien.delahaye@neel.cnrs.fr}
\vspace{10pt}
\begin{indented}
\item[]November 2021
\end{indented}

\begin{abstract}
This paper describes the observation of non-equilibrium field effects at room temperature in four disordered insulating systems: granular Al, discontinuous Au, amorphous NbSi and amorphous InOx thin films. The use of wide enough gate voltage ranges and a cautious analysis of the data allow us to uncover memory dips, the advocated hallmark of the electron glass, in the four systems. These memory dips are found to relax slowly over days of measurements under gate voltage changes, reflecting the impossibility for the systems to reach an equilibrium state within experimentally accessible times. Our findings demonstrate that these electrical glassy effects, so far essentially reported at cryogenic temperatures, actually extend up to room temperature.
\end{abstract}

%
%
%
%
%

\section{Introduction}
\label{intro}

As electrons are very light particles, they are generally associated with very short characteristic times in solids. The fact that slow and glassy phenomena are found in the electronic properties of some disordered insulating systems is thus a challenging and fascinating issue (see \cite{PollakBookElectronGlass2012} for a review). These glassy phenomena are characterized by two main features. First, the electrical conductance is found to decrease slowly after a rapid cooling. This decrease, logarithmic in time, continues at least over weeks of measurements and reflects the impossibility for the system to reach a true equilibrium state. Second, when the disordered insulating systems are thin films used as the (weakly) conducting channel of MOSFET devices, a fast gate voltage ($V_g$) sweep, performed after a long enough time has been spent under a fixed $V_g$ (subsequently noted $V_{geq}$), reveals a conductance minimum centred on this $V_g$ value. This field effect feature, called the ``memory dip'' (MD), is the fingerprint of the gate voltage history experienced by the sample. This MD behaviour and other related features, first investigated in detail in amorphous and microcrystalline indium oxide (InOx) films \cite{OvadyahuPRB1991, OvadyahuPRB1993}, have since then proven to be rather common among disordered insulating systems. Indeed, they were also observed in discontinuous and ultrathin continuous films of metals \cite{AdkinsJPC1984, GoldmanPRL1997, FrydmanEL2012}, in granular Al \cite{GrenetEPJB2003, GrenetEPJB2007}, in oxidized Be \cite{OvadyahuPRB2010}, in microcrystalline $Tl_2O_{3-x}$ \cite{OvadyahuPRB2013}, in amorphous and microcrystalline metal-semiconductor alloys (NbSi \cite{DelahayeEL2014, DelahayeSciPost2020}, GeSbTe \cite{OvadyahuPRB2015}, GeTe \cite{OvadyahuPRB2016}, GeBiTe \cite{OvadyahuPRB2018a}), with the notable exception of standard doped semi-conductors. It was soon suggested that these glassy features are the signature of the electron glass \cite{PollakBookElectronGlass2012}. This glassy state, theoretically predicted in the 80ies, stems from the coexistence of disorder and ill screened electron-electron interactions in disordered systems \cite{DaviesPRL1982, GrunewaldJPCS1982, PollakSEM1982}. Recently it was found that the upper bound of the relaxation time distribution decreases and becomes measurable when both the charge carrier concentration and resistance per square are diminished in amorphous InOx films \cite{OvadyahuPRB2018b}, strongly supporting the electron glass hypothesis.

Interestingly, most of the experimental works devoted to these glassy effects have been limited so far to the liquid helium temperature ($T$) range. At 4~K, MDs represent typically a few \% of the conductance when the sheet resistance $R_{s4K}$ lies in the range $1~M\Omega - 1~G\Omega$. In the few studies where higher $T$ were explored (up to about 50~K), the relative amplitudes of the MDs were found to decrease strongly when $T$ increases (as a power law in granular Al films \cite{GrenetEPJB2007}, exponentially in InOx films \cite{OvadyahuEL1998}). In InOx films, such a decrease led to the conclusion that glassy effects are indeed absent or too small to be measurable above a few tens of kelvins \cite{OvadyahuPRB2002}. But two experimental findings have recently called the generality of this conclusion into question. First, MDs of the order of 1~\% were reported at room $T$ in insulating discontinuous Au films \cite{FrydmanPRL2016, EisenbachPhD2017}. Second, although very small (less than 0.02~\%), traces of MDs were found to persist up to room $T$ in an amorphous NbSi thin film \cite{DelahayeEL2014}.

Whether room $T$ glassy effects are exceptional or widespread is the question we would like to answer in this paper. In a first part, we present field effect measurements performed at room $T$ on insulating granular Al films covering a large resistance range. MDs that respond slowly to $V_{geq}$ changes are observed in all the films studied. Their amplitudes can be as large as a few percent if the resistance of the film is high enough. These results are compared to those obtained on discontinuous Au and on amorphous NbSi. We then demonstrate that MDs are also present in InOx films at room $T$, although they are mixed with another conductance relaxation of different nature. We end by discussing why room $T$ glassy effects have been so scarce in previous studies, and what are the new perspectives offered by our findings.

\section{Samples and measurement set-up}
\label{method}

The granular Al films measured in this study are 10~nm-thick. They were made by the electron beam deposition of pure Al in a partial pressure of $O_2$ \cite{GrenetEPJB2007}. The Al evaporation rate was $1.8~\AA/s$ and a large range of room $T$ sheet resistances was obtained by changing the $O_2$ partial pressure, from $R_{s300K} \simeq 100~k\Omega$ ($P_{O2} = 2.4\times 10^{-5}~mbar$) up to $100~G\Omega$ ($P_{O2} = 3.4\times 10^{-5}~mbar$).

The amorphous NbSi film is 2.5~nm thick and has a Nb content of 13~\%. This sample, already measured at 4.2~K in previous works \cite{DelahayeEL2014, DelahayeSciPost2020} was made by the co-deposition of Nb and Si at room $T$ under high vacuum (typically a few $10^{-8}~mbar$) \cite{CraustePRB2013}. Its sheet resistance at room $T$ is $25~k\Omega$ ($100~M\Omega$ at 4.2~K).

\begin{figure}[!h]
\centering
\includegraphics[width=10cm]{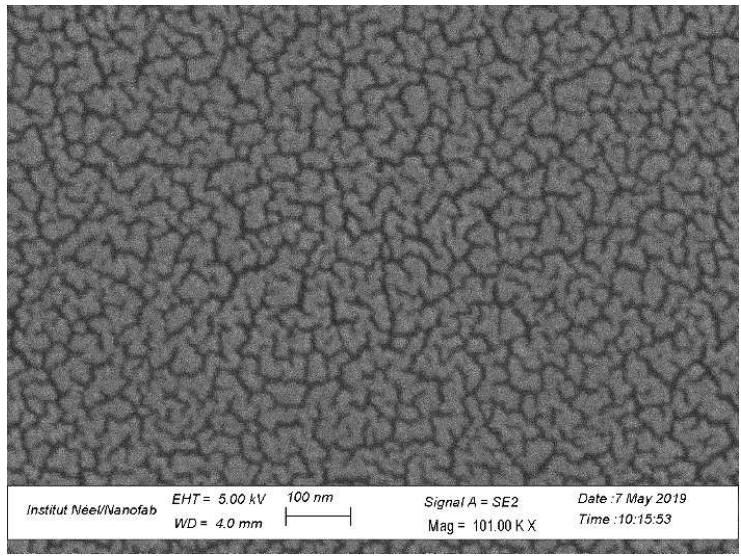}
\caption{SEM picture of a discontinuous Au film evaporated at room $T$ on a $Si_{++}/SiO_2$ substrate. Nominal thickness: 7.0~nm;  $R_{s300K} = 300~M\Omega$.}\label{Figure0}
\end{figure}

The insulating discontinuous Au films were made by the electron beam deposition of Au at room $T$. The evaporation rate was varied between 0.2 and $2~\AA/s$, under a residual vacuum of $10^{-6}$ to $10^{-7}$ mbar. The films have nominal thicknesses of 6 - 7~nm and their SEM pictures reveal a characteristic maze structure (see Fig.~\ref{Figure0}). Their resistance was found to increase after deposition, as is commonly observed in insulating discontinuous Au films \cite{AndersonJAP1976}, and becomes roughly stable after a few days of measurements. The films used in this study have relaxed $R_{s300K}$ values between $30~M\Omega$ and $30~G\Omega$. 20~nm-thick gold contacts were evaporated prior to the discontinuous Au film.

We also show results obtained on a 10~nm-thick amorphous InOx film made by the electron beam deposition of $In_2O_3$ at a rate of $1~\AA/s$, under a partial $O_2$ pressure of $2\times 10^{-4}~mbar$, having $R_{s300K} \simeq 300~k\Omega$.

In order to perform field effect measurements, the disordered insulating thin films were all deposited on highly doped Si wafers ($Si_{++}$, the gate) coated with 100~nm of thermally grown $SiO_2$ (the gate insulator). The active channel has a typical size of $1~mm \times 1~mm$ and its electrical conductance $G$ was measured by using a two-terminal DC technique. For room $T$ measurements ($T$ between 293~-~296~K), a bias voltage of about 1~V was applied with a low noise voltage source, and the current was measured with the Femto DLPCA 200 current amplifier. Unless otherwise mentioned, the device charging currents induced by $V_g$ changes were negligible compared to the bias current variations related to the slow processes discussed in this paper (this was checked by performing $V_g$ sweeps with zero bias, which allow the measurement of charging currents alone). $V_g$ could be varied between -30~V and 30~V, far enough from the practical breakdown limit of our 100~nm-thick $SiO_2$ insulating barrier which is around 50~V. No significant leaking current was detected within this $V_g$ range. All the samples were mounted in a metallic enclosure and maintained in vacuum during the measurement campaigns.

\section{Evidence for room temperature memory dips in granular Al films}

\begin{figure}[!h]
\centering
\includegraphics[width=16cm]{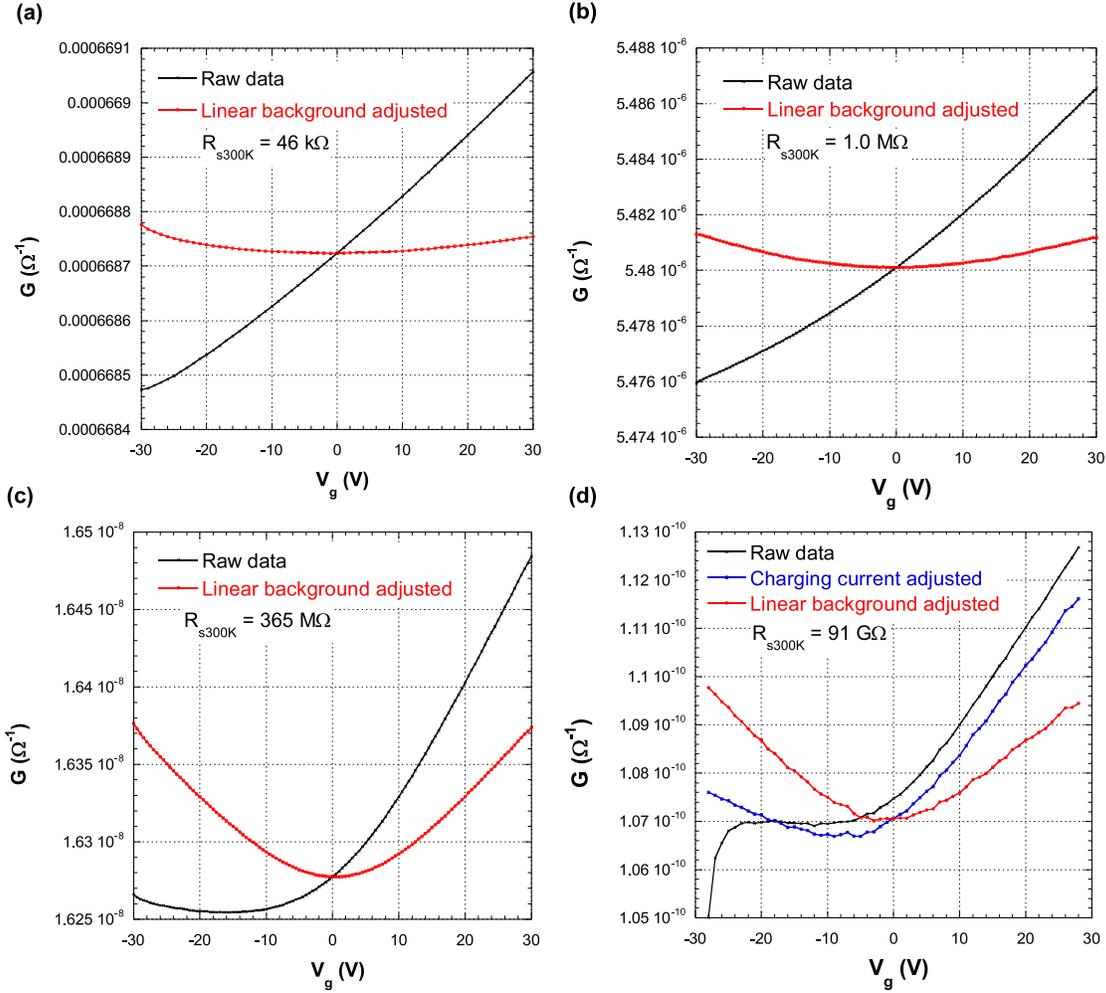}
\caption{$G(V_g)$ curves measured at room $T$ after about one day under $V_{geq} = 0~V$ in four different granular Al films. Black symbols: raw data; blue symbols: charging current adjusted data (this correction is necessary only when $R \geq 1~G\Omega$); red symbols: linear background adjusted data (refer to text for details). (a) $R_{s300K} = 46~k\Omega$; (b) $R_{s300K} = 1.0~M\Omega$; (c) $R_{s300K} = 365~M\Omega$; (d) $R_{s300K} = 91~G\Omega$.}\label{Figure1}
\end{figure}

We first sought to find out whether electron glass features are present at room $T$ in granular Al films. At 4~K, their hallmark is the existence of a MD centred on the cool-down $V_g$ \cite{GrenetJPCM2017} and the slow formation of a new one if $V_g$ is set to a new value \cite{GrenetEPJB2003, GrenetEPJB2007}. At room $T$, we first simply examined whether a MD exists centred on $V_g = V_{geq}$ applied for long enough. The protocol is thus the following: $V_g$ is fixed to 0~V for a long enough time (one day or more) before fast $V_g$ sweeps are performed around this value. The $V_g$ sweeps between -30~V and 30~V are typically 200~s long and can be repeated every 2000 - 4000~s in order to average and reduce the noise level of the curves (these parameters can differ by a factor of 2 depending on the sample). The results (black curves) are displayed in Fig.~\ref{Figure1} for four different films having $R_{s300K}$ values between $46~k\Omega$ and $91~G\Omega$. The less resistive film with $R_{s300K} = 46~k\Omega$ has a 4.2~K resistance of $8~M\Omega$, typical of the samples usually measured in liquid He.

The $G(V_g)$, even when corrected for transitory charging effects (significant only when $R_s$ is above $1~G\Omega$), reveal either no conductance minimum (samples with $R_{s300K} = 46~k\Omega$ and $1~M\Omega$) or a minimum which is not centred on $V_{geq}$ (samples with $R_{s300K} = 365~M\Omega$ and $91~G\Omega$). However, as we shall demonstrate below, MDs are indeed present in all the samples, although they do not show here under their usual 4~K form. First, they coexist with a comparatively large ``normal'' linear field effect which gives rise to a monotonic increase of the conductance with $V_g$. This linear field effect contribution is usually attributed to the energy dependence of the thermodynamic density-of-states at the Fermi level \cite{OvadyahuPRB2002}. Second, the MDs extend over a $V_g$ range larger than our accessible $V_g$ window (-30~V to 30~V) and only the bottom of the dips is visible.

\begin{figure}[!h]
\centering
\includegraphics[width=10cm]{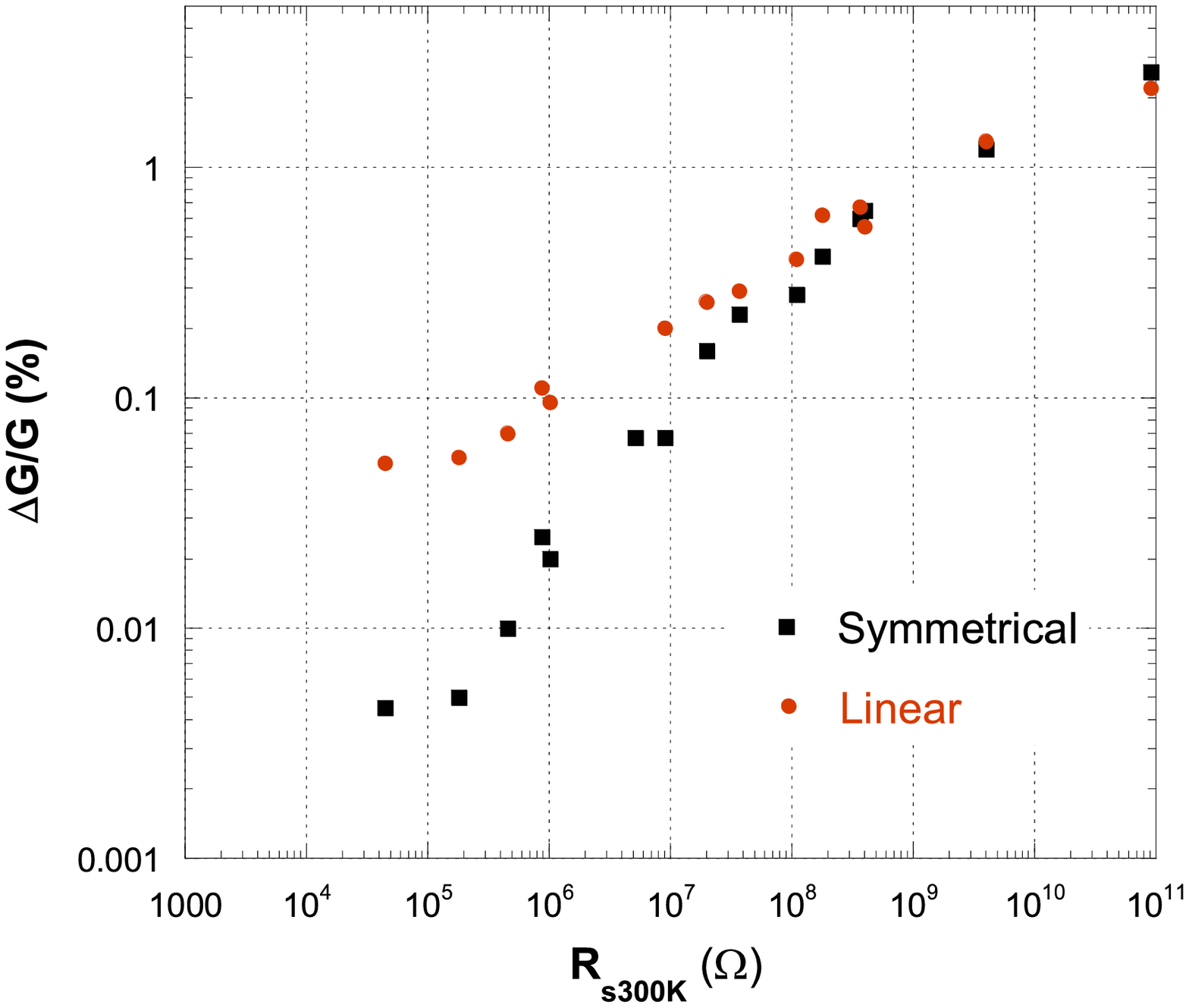}
\caption{Evolution of the symmetrical (black square symbols) and linear (red dot symbols) field effect components with $R_{s300K}$ in different granular Al films. Their relative amplitudes were calculated from $G(V_g)$ curves between 0~V and 30~V (see the text for details).}\label{Figure2}
\end{figure}

If we subtract a linear contribution from each $G(V_g)$ curves so as to equalize the conductance at -20~V and 20~V, we get conductance minima (red curves in Fig. 1) that are roughly symmetrical around $V_{geq} = 0~V$. The conductance change between $V_{geq}=0~V$ and $V_g$ can thus be decomposed in two parts : a linear one, $\Delta G_{lin}$, deduced from the $V_g$ slope of the linear contribution, and a (roughly) symmetrical one, $\Delta G = G(V_g) - G(V_{geq}=0~V)- \Delta G_{lin}(V_g)$. The relative amplitudes $\Delta G_{lin}/G$ and $\Delta G/G$ of respectively the linear and symmetrical contributions measured between 0~V and 30~V both decrease as $R_s$ is lowered (see Fig.~\ref{Figure2}). They are roughly the same as long as $R_s > 1~G\Omega$, but, below this resistance range, the symmetrical component decreases faster than the linear one, which explains why no conductance minima is discernable in low $R_s$ films.

\begin{figure}[!h]
\centering
\includegraphics[width=10cm]{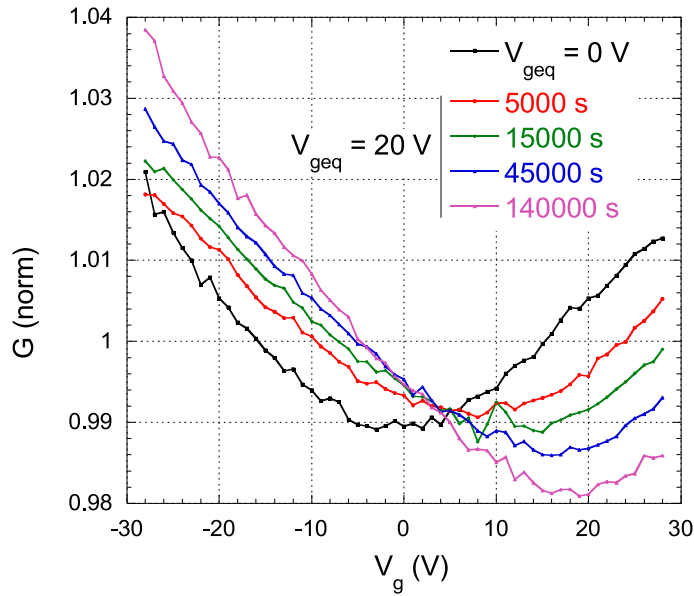}
\caption{Normalized $G(V_g)$ curves measured at room $T$ before (black curve) and at different times after $V_{geq}$ was changed from 0~V to 20~V (the same linear background has been subtracted from all the raw curves). The sample, a granular Al film with $R_{s300K} = 91~G\Omega$, was first kept for a few days under $V_{geq} = 0~V$.}\label{Figure3}
\end{figure}

In order to prove that these conductance minima are MDs, we still need to verify that they evolve after a $V_{geq}$ change. This is shown in Fig.~\ref{Figure3} where $G(V_g)$ curves have been measured at different times after $V_{geq}$ was changed from 0~V to 20~V. The $V_{geq} = 0~V$ linear background correction is assumed to remain the same in a given sample when $V_{geq}$ is changed. The conductance minimum, first centred on 0~V, clearly moves toward the new $V_{geq}$ value. The fact that it is not centred on the new $V_{geq}$ value for short times can simply be explained by the strong overlap between the new and the old dips (see the discussion below).

\begin{figure}[!h]
\centering
\includegraphics[width=16cm]{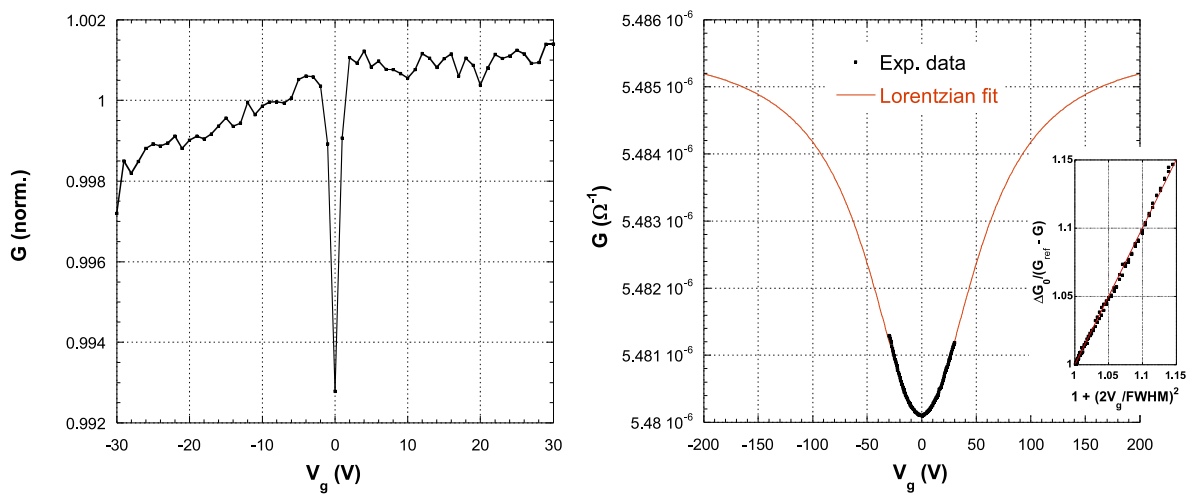}
\caption{Left panel: normalized $G(V_g)$ curves measured 16800~s after a cooling from room $T$ to 4.2~K in a granular Al film ($R_{s300K} = 46~k\Omega$, $R_{s4.2K} = 8.0~M\Omega$). Right panel: $G(V_g)$ curve measured at room $T$ (black symbols) and its Lorentzian fit $G = G_{ref} - \Delta G_0 /(1+(2V_g/FWHM)^2)$ (red line, $G_{ref} = 5.5\times 10^{-6}~\Omega^{-1}$, $\Delta G_0 = 5.5\times 10^{-9}~\Omega^{-1}$, $FWHM = 120~V$) in a granular Al film having $R_{s300K} = 1.0~M\Omega$. Insert: $\Delta G_0/(G_{ref}-G)$ as a function of $1+ (2V_g/FWHM)^2$ between -20~V and 20~V. The uncertainty about the FWHM value is of $\pm 20~V$.}\label{Figure4}
\end{figure}

We now discuss the MD widths. For comparison, the less resistive film of our series ($R_{s300K} = 46~k\Omega$) was measured at 4.2~K on the same $V_g$ range (left panel of Fig.~\ref{Figure4}). At 4.2 K, the full MD is clearly visible and much thinner than our $V_g$ window, with a full width at half maximum (FWHM) of about only 1.5~V, in agreement with previous studies \cite{DelahayePRL2011}. The $T$ evolution of the MD shape was studied previously between 4~K and 20~K in granular Al films \cite{GrenetEPJB2007}. It was found that, except at the lowest $T$, the MDs are well described by a Lorentzian shape and that their typical widths grow roughly linearly with $T$. Assuming that this law remains valid up to room $T$ and that the 4.2~K width is roughly $R_s$ independent \cite{DelahayePRL2011}, we expect a width of about 100~V at 300~K. By fitting the red curves of Fig.~\ref{Figure1} with Lorentzian functions centred on 0~V (the equilibrium $V_g$), we get FWHM estimates of 150~V, 120~V, 90~V and 70~V for granular Al films of respectively $R_s = 46~k\Omega$, $1.0~M\Omega$, $365~M\Omega$ and $91~G\Omega$, in rough agreement with our expectation. We observed a similar decrease in the FWHM values as $R_s$ increases at 4~K but with no explanation so far. Such a fit (red line) and the measured $G(V_g)$ curve (black symbols) are plotted in the right panel of Fig.~\ref{Figure4} for the $R_{s300K} = 1.0~M\Omega$ film. It thus appears clearly that for all films, our $V_g$ window of 60~V allows us to only measure the bottom of the MD, most of it being outside this $V_g$ range. Such a result stresses an important practical point: in order to be able to measure room $T$ glassy effects in granular Al films, the $V_g$ range investigated should be large enough. We shall see below that this conclusion applies to other systems as well.

\begin{figure}[!h]
\centering
\includegraphics[width=10cm]{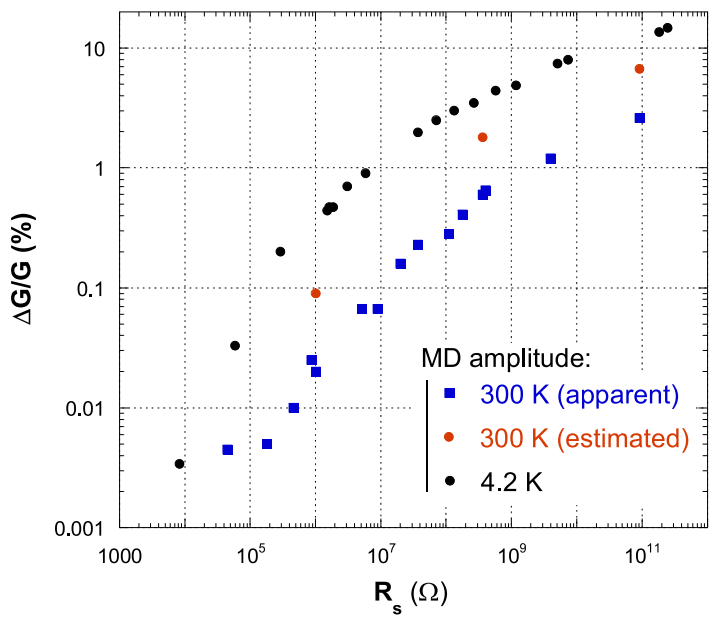}
\caption{MD amplitude as a function of $R_s$ in granular Al films. Black symbols: 4.2~K values as a function of $R_{s4K}$ (the full MDs are visible at this temperature, data from Ref. \cite{DelahayePRL2011}). Red and blue symbols : room $T$ values as a function of $R_{s300K}$. Red symbols: the room $T$ amplitudes of the MD are estimated from Lorentzian fits of the data between -30~V and 30~V ($\Delta G$ is given by the parameter $\Delta G_0$ of the Lorentzian fit). Blue symbols: the room $T$ amplitudes are measured between 0~V and 30~V ($\Delta G = \Delta G(30~V)$).}\label{Figure5}
\end{figure}

The relative amplitudes of the MDs are given by the symmetrical component values $\Delta G/G$, when $|V_g-V_{geq}|$ is large enough so that $\Delta G$ saturates. We can get lower bound values by comparing the conductance at $V_g = V_{geq} = 0~V$ and at $V_g = 30~V$ after one day or more under $V_{geq}$, that is by measuring $\Delta G(30V)/G$. A clear increase of $\Delta G(30V)/G$ is observed with $R_s$, as seen in previous 4.2~K studies \cite{DelahayePRL2011}: this lower bound value reaches 2.5~\% when $R_{s300K} \sim 100~G\Omega$, and is still measurable although very small ($4.5 \times 10^{-3}~\%$) when $R_{s300K} \sim 100~k\Omega$ (blue symbols in Fig.~\ref{Figure5}). According to our Lorentzian fits, the actual values of the full MDs ($\Delta G_0/G$) are 2.5 and 4 times larger for $R_s$ around respectively $100~G\Omega$ and $100~k\Omega$. These relative MD amplitudes are only a few times smaller than the ones measured at 4.2~K on granular Al films having similar $R_s$ values (see Fig.~\ref{Figure5}). Note that since the MD amplitudes are believed to depend only logarithmically on the sweep parameters and the time spent under $V_{geq}$ (see the dynamics tests below), the comparison between the different samples highlighted in Fig.~\ref{Figure5} is meaningful although these parameters were not strictly the same for the different samples.

\begin{figure}[!h]
\centering
\includegraphics[width=16cm]{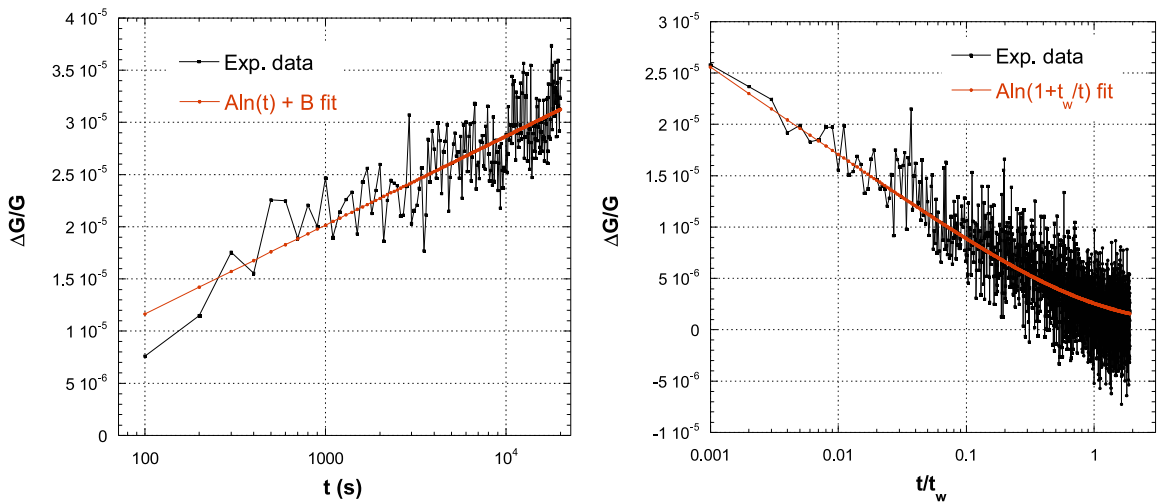}
\caption{Left panel: time evolution of $\Delta G/G = [G(-20~V)-G(20~V)]/G(-20~V)$ reflecting the growth of the new MD at 20~V after $V_{geq}$ was changed from 0 to 20~V. Right panel: erasure of the same 20~V MD (formed during $t_w = 20000~s$) after $V_{geq}$ is switched back to 0~V. The linear field effect contribution was removed to show the MD amplitude going asymptotically to zero in the right panel. Granular Al film, $R_{s300K} = 46~k\Omega$. }\label{Figure6}
\end{figure}

To further characterize the conductance dynamics associated with $V_{geq}$ changes, we applied the isothermal 2 dips protocol, usually used at 4~K \cite{OvadyahuPRB2000, OvadyahuPRB2002, GrenetEPJB2007}: after a long stay of one day or more under $V_{geq1} = 0~V$, $V_{geq}$ is set to a new value $V_{geq2} = 20~V$ for a given time $t_w$ (step 1), and then shifted back to $V_{geq1}$ (step 2). The time evolution of the conductance at -20~V (ref. value), 0~V and 20~V are recorded during step 1 and step 2. Typical results for the less resistive granular Al sample (hence the high noise level of the curves) are plotted in Fig.~\ref{Figure6}. The difference between the conductance value at -20~V (ref.) and at 20~V ($V_{geq2}$) increases as a logarithm of the time elapsed during step 1 (new MD formation), and follows the erasure function $A\ln(1 + t_w/t)$ during step 2 (new MD erasure), as is commonly observed at 4.2~K in granular Al and InOx films \cite{GrenetEPJB2007, AmirARCMP2011}. This shows that also at room $T$, the conductance relaxations are associated to log-uniformly distributed relaxation times, at least in the time window explored here ($\sim 100~s$ to $10^5~s$), with modes effectively relaxing back and forth under $V_{geq}$ changes. Similar logarithmic relaxations are observed at room $T$ in all our granular Al films.

\section{Comparison with amorphous NbSi and discontinuous Au films}

\begin{figure}[!h]
\centering
\includegraphics[width=16cm]{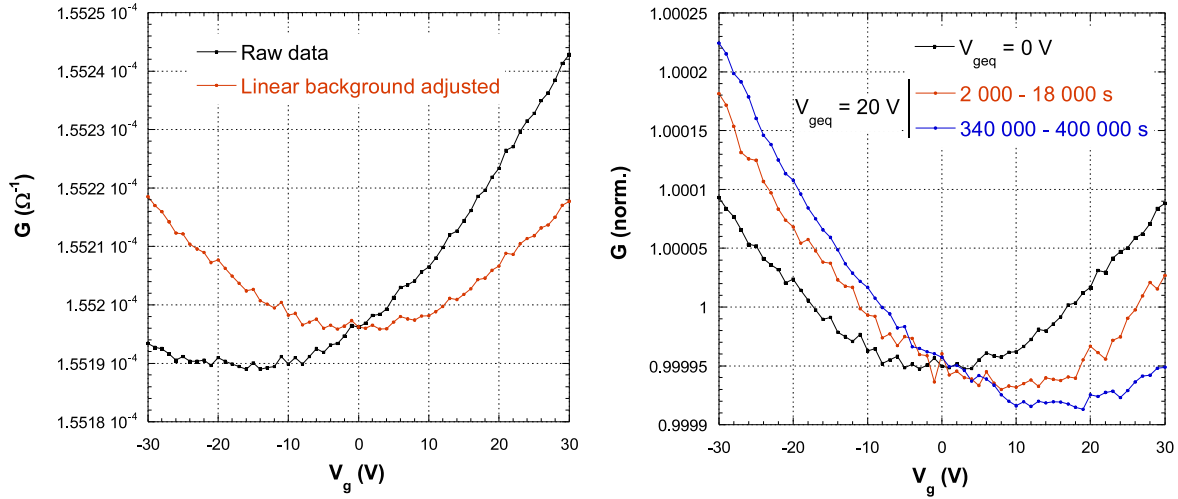}
\caption{Left panel: $G(V_g)$ curve measured at room $T$ after 20 days spent under $V_{geq} = 0~V$. Black symbols: raw data; red symbols: linear background adjusted data. Right panel:  evolution of the $G(V_g)$ curves (corrected for the linear background) when $V_{geq}$ is changed from 0~V to 20~V. The $G(V_g)$ curves have been averaged over the time intervals indicated in the legend in order to improve the signal-to noise ratio. Amorphous $Nb_{13}Si_{87}$ film, 2.5~nm thick, $R_{s300K} = 25 ~k\Omega$.}\label{Figure7}
\end{figure}

We will now see how granular Al compares with the two other systems in which room $T$ glassy effects have been previously reported, i.e. amorphous NbSi and discontinuous Au films.
In the amorphous NbSi film measured in Refs. \cite{DelahayeEL2014, DelahayeSciPost2020} ($R_{s300K} = 25~k\Omega$), a symmetrical conductance minimum centred on the equilibrium gate voltage value is visible at room $T$ after the subtraction of a linear field effect contribution (left panel of Fig.~\ref{Figure7}). This conductance minimum moves slowly after a $V_{geq}$ change (right panel of Fig.~\ref{Figure7}) and is therefore a MD. Its amplitude between 0~V and 30~V is of 0.015~\%, slightly higher than what is found in granular Al films of similar $R_s$ values. Like in granular Al films, only the bottom of the MD is visible in our $V_g$ window: a fit with a Lorentzian shape gives a full width at half maximum of 140~V ($\pm 20~V$) and a full amplitude of 0.09~\%. MD drifts are observed over days after $V_{geq}$ changes, which is a clear indication that very long relaxation times are still involved in the room $T$ dynamics.

\begin{figure}[!h]
\centering
\includegraphics[width=16cm]{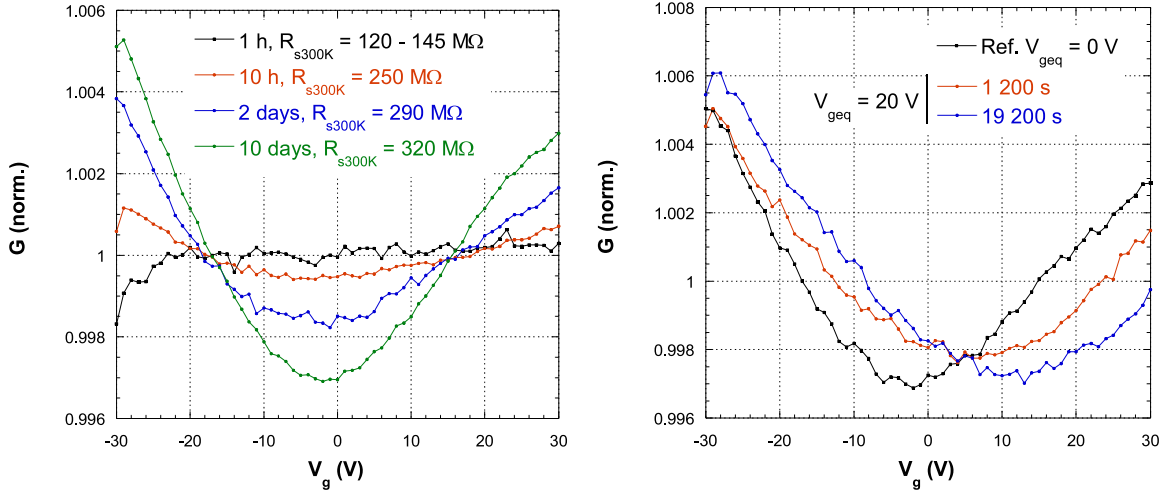}
\caption{Left panel: evolution of the normalized $G(V_g)$ curves (corrected for a linear background) for a discontinuous Au film as a function of the time elapsed at room $T$ since its fabrication. The $R_{s300K}$ value also increases during this time. Right panel: $G(V_g)$ evolution resulting from a $V_{geq}$ change from 0~V to 20~V.}\label{Figure8}
\end{figure}

In discontinuous Au films made at room $T$, MDs are present but with some specificities. First, the linear field effect has a negative slope, i.e. the conductance decreases slightly with $V_g$, a feature already mentioned in some other systems (Be \cite{OvadyahuPRB2010}, GeSbTe \cite{OvadyahuPRB2015}, GeTe \cite{OvadyahuPRB2016}, GeBiTe \cite{OvadyahuPRB2018a}). It was suggested that the sign of this slope is controlled by the energy derivative of the thermodynamic density-of-states at the Fermi level \cite{OvadyahuPRB2010}. Second, shortly after the film fabrication (typically a few hours), MDs are not seen. In the example of Fig.~\ref{Figure8}, a conductance minimum becomes clearly visible only after a few days at room $T$, when the resistance value has stabilised (in the example of Fig.~\ref{Figure8}, it reaches about 1~\% between 0~V and 30~V). This last feature, which to our knowledge was not reported before, deserves further investigation. It may reflect a relationship between the resistance relaxation which follows the film fabrication and which is usually attributed to the evolution of Au grain shapes \cite{AndersonJAP1976}, and the MD formation. However, like in granular Al and amorphous NbSi films, $V_{geq}$ changes lead to logarithmic-like relaxations of the conductance. The full MDs are clearly not visible in our $V_g$ range, but since the curves are not strictly symmetrical, the actual MD width could not be estimated with precision. The amplitude of the MD observed between 0~V and 30~V grows from 0.3~\% for $R_{s300K} = 30~M\Omega$ to 1~\% for $R_{s300K} = 15~G\Omega$.

\section{Room temperature electron glassiness in an amorphous InOx film}

\begin{figure}[!h]
\centering
\includegraphics[width=10cm]{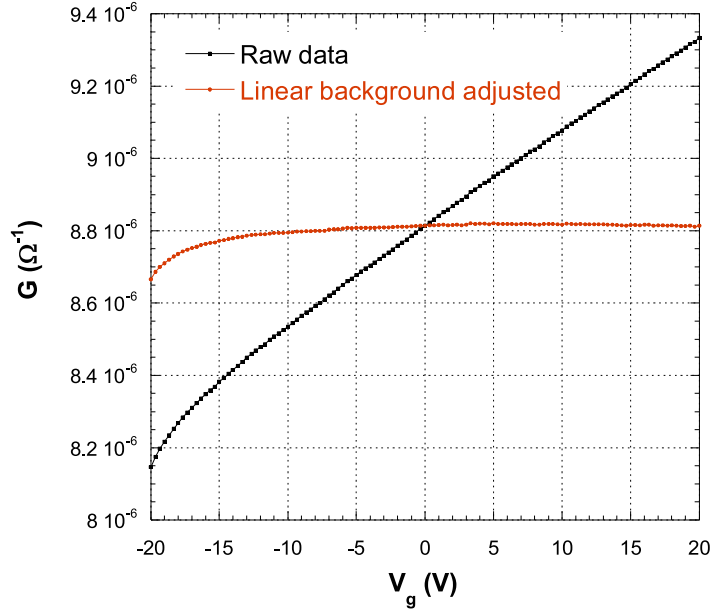}
\caption{$G(V_g)$ curve measured at room $T$ after 3 hours spent under $V_{geq} = 10~V$. Black symbols: raw data; red symbols: linear background adjusted data (equalizing $G(0~V)$ and $G(20~V)$). Amorphous InOx film, $R_{s300K}\simeq 300~k\Omega$.}\label{Figure9}
\end{figure}

Insulating InOx is the archetypal electron glass system and its study at room $T$ is thus of prime importance. But as we will now show, the identification of MDs and their relaxations turns out to be trickier than in the other systems.
By using the same protocol as before, i.e. $G(V_g)$ curves are measured after a long enough stay under a given $V_{geq}$ value (10~V in Fig.~\ref{Figure9}), no conductance minimum is visible even if we remove a linear field effect contribution. Instead, a fast increase of the conductance is observed around -20~V, i.e. the starting side of the $V_g$ sweep, followed by a soft concave curvature at higher $V_g$ values. The $G(V_g)$ curve can thus definitely not be interpreted as the mere sum of a linear normal field effect and a symmetrical MD centred on $V_{geq}$. Note also the large amplitude of the linear field effect contribution, which represents a 6~\% change of $G$ between 0~V and 20~V.

\begin{figure}[!h]
\centering
\includegraphics[width=16cm]{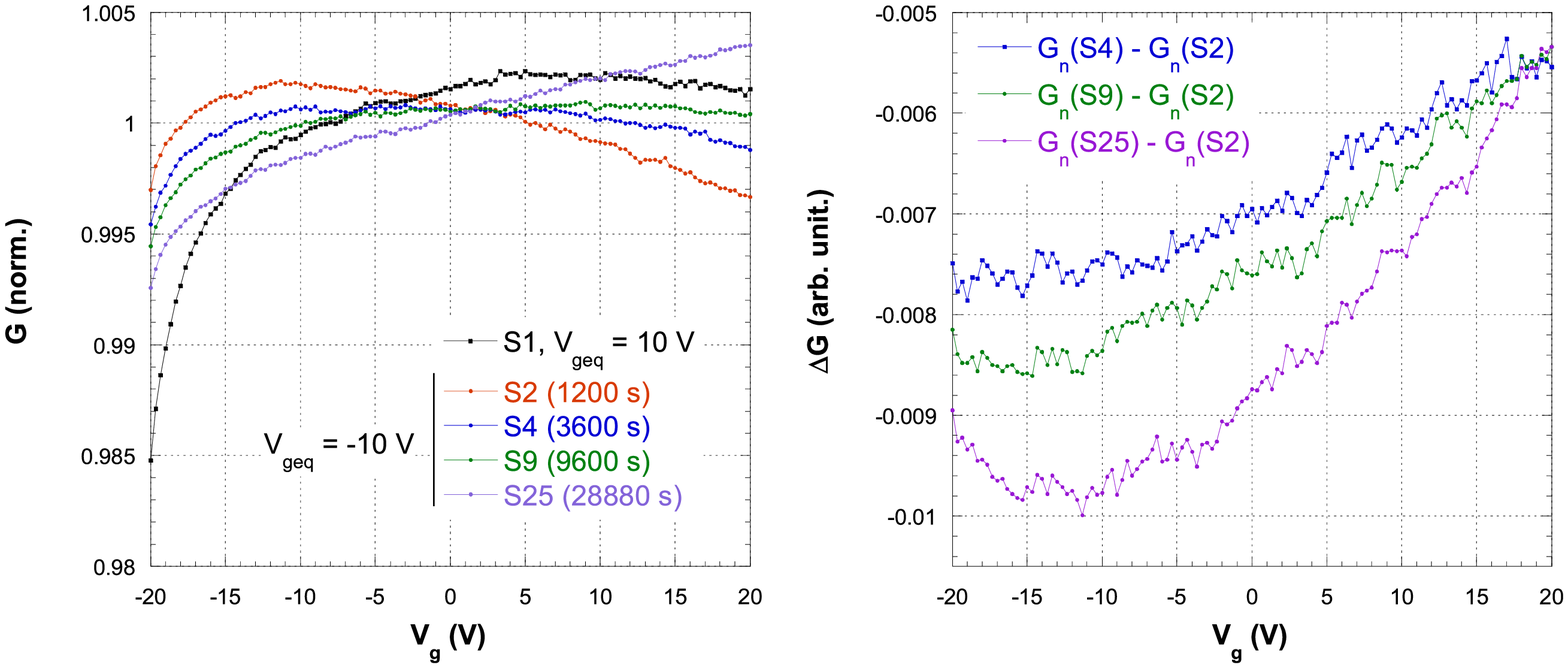}
\caption{Left panel: time evolution of normalized $G(V_g)$ curves corrected for the linear background after $V_{geq}$ is changed from 10~V to -10~V. Sn stands for ``nth $V_g$ sweep''. S1 corresponds to the curve of Fig.~\ref{Figure9}, taken before the $V_{geq}$ change. Right panel: differences between normalized $G(V_g)$ curves and S2 (vertically shifted in order to equalize their 20~V values). $V_g$ sweeps are performed from left to right. Amorphous InOx film, $R_{s300K} \simeq 300~k\Omega$.}\label{Figure10}
\end{figure}

If we now change $V_{geq}$ from 10~V to -10~V, the time evolution of the normalized, linear background corrected curves is also quite startling (Fig.~\ref{Figure10}). Between sweep No 2 (curve S2, taken 1200~s after the change to -10~V) and the next ones, the normalized conductance is found to decrease continuously with time for negative $V_g$, and to increase for positive $V_g$ (left panel of Fig.~\ref{Figure10}). A direct plot of the difference between sweep 2 and the following ones indeed suggests that the $V_{geq}$ change provokes the formation of a wide MD centred on -10~V (right panel of Fig.~\ref{Figure10}). Its height between -10~V and 20~V amounts to 0.4~\% of the conductance, which is larger than what we get in granular Al film having similar $R_s$ values. Between sweep No 1 (curve S1, taken before the $V_{geq}$ change) and sweep No 2 (curve S2), the conductance changes the opposite way, with an increase for negative $V_g$ and a decrease for positive $V_g$.

\begin{figure}[!h]
\centering
\includegraphics[width=16cm]{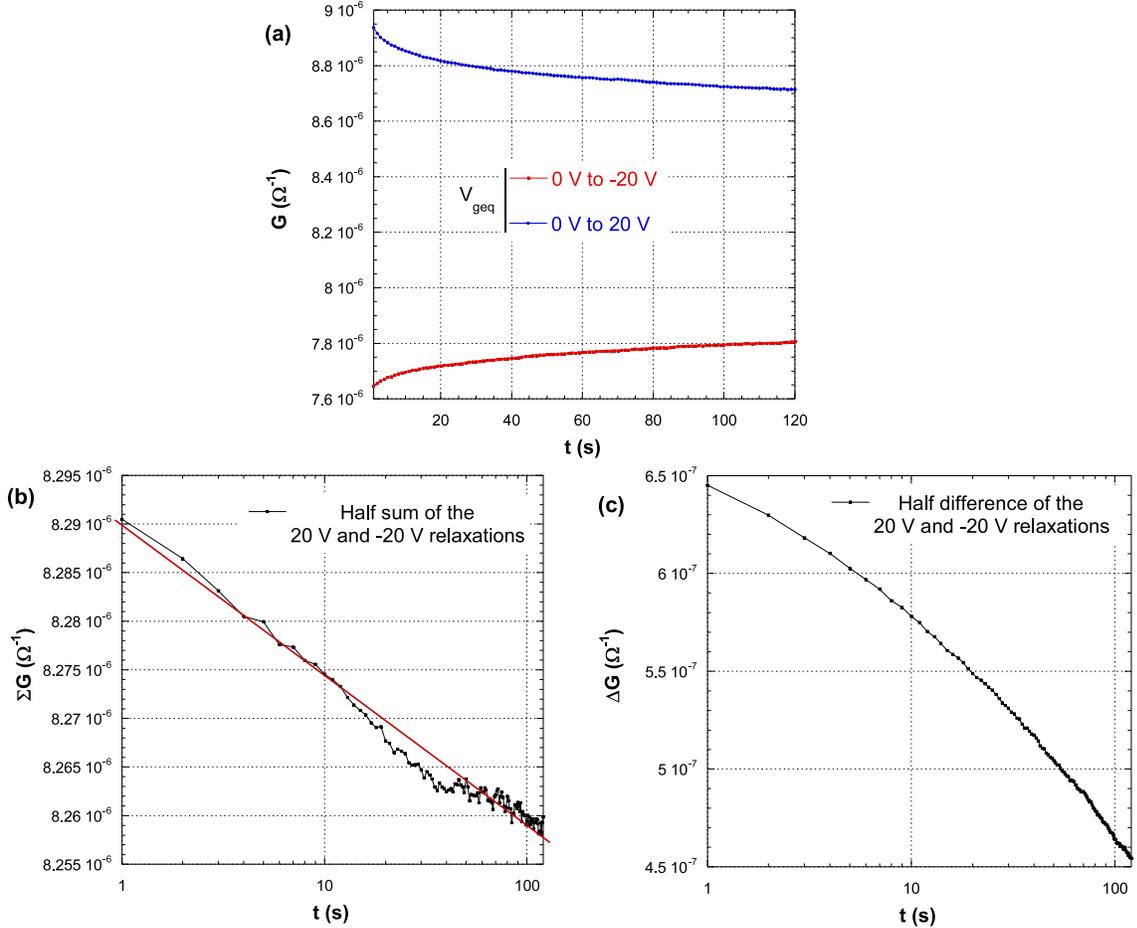}
\caption{(a) Conductance relaxations measured after $V_{geq}$ is changed from 0 to 20~V and from 0 to -20~V (see text for details). (b) Half sum of the 20~V and -20~V relaxations, corresponding to the ``usual'' MD formation (the red-line is a $\log(t)$ fit). (c) Half difference of the 20~V and -20~V relaxations, corresponding to an anti-symmetrical relaxation. Amorphous InOx film, $R_{s300K} \simeq 300~k\Omega$.}\label{Figure11}
\end{figure}

One way to interpret this behaviour is to suppose that two sources of conductance relaxation are present: one which dominates the observed evolution between the two first scans S1 and S2, and the memory dip formation which dominates after the second scan S2.  This hypothesis is supported by the following experiment. After a long stay under 0~V, we change the gate voltage to a new value and measure the time evolution of the conductance at that value (no $V_g$ sweep). We compare what happens when $V_g$ is switched from 0 to 20~V and from 0 to -20~V. If only the usual memory dip formation was present, the two relaxations would be roughly the same and consist in a slow $\ln(t)$ decrease of the conductance. However we observe two relaxations of opposite signs: a conductance increase if $V_g$ is changed from 0~V to -20~V and a conductance decrease if it is changed from 0~V to 20~V. These are shown in Fig.~\ref{Figure11}a. It is seen that the two conductance relaxations are not exactly opposite. By taking their half sum and difference, it is possible to isolate a dominant contribution which is anti-symmetrical with respect to the sign of the $V_g$ jump, and a symmetrical one consisting of the usual downward logarithmic relaxation associated to the formation of a MD (see Fig.~\ref{Figure11}b  and \ref{Figure11}c). Note that the anti-symmetrical relaxation measured for 100~s is seven times larger than the one associated to the MD.

The anti-symmetrical contribution explains qualitatively the anomalous S1 curve observed in Fig. 10. Just before sweep S1 started, $V_g$ was changed from 10~V to -20~V. This -30~V jump caused an upward conductance drift during sweep S1, which mostly distorts the left part of the curve. Sweep S2 and the subsequent ones are preceded by a $V_g$ jump three times smaller (from -10~V to -20~V). It gives rise to a smaller distortion of the curves, which can be eliminated by subtracting S2 from the subsequent curves. Hence the clear demonstration of the MD growth shown in Fig.~\ref{Figure10} right panel.

\begin{figure}[!h]
\centering
\includegraphics[width=16cm]{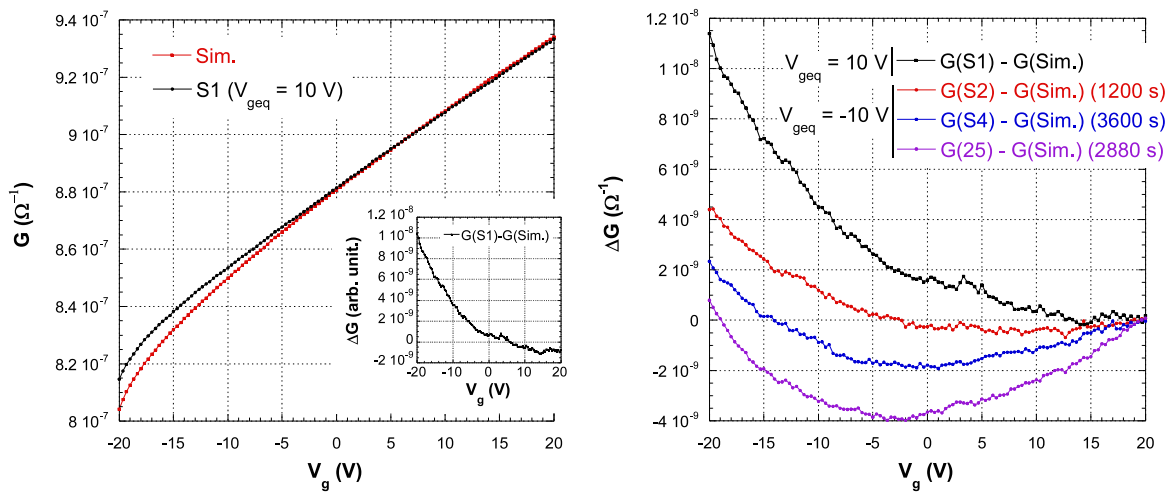}
\caption{Left panel: $G(V_g)$ curve of Fig.~\ref{Figure9} and \ref{Figure10} (S1) measured after 3 hours under $V_{geq} = 10~V$ (black symbols). The simulated curve takes into account the anti-symmetrical relaxation (red symbols, see text for details). Insert: difference between the measured $G(V_g)$ curve (S1) and the simulated one. Right panel: time evolution of the difference between the measured $G(V_g)$ curves and the simulated ones after $V_{geq}$ is changed from 10~V to -10~V. The progressive left shift of the curves minimum is created by the strong overlap of the fading 10~V MD and the growing -10~V one.}\label{Figure12}
\end{figure}

We may be more quantitative and try to simulate the anti-symmetrical contribution to the curves of Fig.~\ref{Figure9} and \ref{Figure10}. In order to do so, we use the curve in Fig.~\ref{Figure11}c to represent the relaxation caused by a $\Delta V_g= 20~V$ jump, and assume that for other jumps the relaxation is simply proportional to $\Delta V_g$. The response for each $V_g$ sweep is then simulated as the sum of the response to the initial negative $V_g$ jump (start of the sweep, from $V_{geq}$ to -20~V in that case) and to the successive small positive jumps composing the sweeps from -20~V to 20~V). Note that the conductance shift caused by the ``normal'' linear field effect is present in the response of Fig.~\ref{Figure11}c and is thus automatically taken into account in the simulation. An offset is added to make the simulated and actual curves coincide at $V_g = 20~V$ (see Fig.~\ref{Figure12}-left).
The difference between the measured and simulated conductance sweeps is expected to represent the MD relaxation contribution. As shown in Fig.~\ref{Figure12}-right, they indeed correspond to a MD around 10~V being progressively erased as a new one centred on -10~V is growing. The progressive shift to the left of the curves minimum is created by the strong overlap of the two MDs.

The precise origin of the anti-symmetrical relaxation is still unclear (no significant charging current is measured when the measurement of Fig.~\ref{Figure11} is reproduced with a zero bias voltage). Interestingly enough, its relative amplitude decreases when $T$ is lowered, which is contrary to the MD contribution. It thus becomes negligible compared to the latter at cryogenic temperatures. This $T$ dependence suggests that the anti-symmetrical relaxation may be related to the movement of ionic species in the films, possibly oxygen vacancies. Conductance relaxations induced by the drifts of oxygen vacancies under electric fields have been observed at room $T$ in $SrTiO_3$ materials \cite{DelahayeJPD2016}.

\section{Why have room temperature glassy effects not been seen before?}

Glassy effects are present in all the systems and all the samples we have been measuring: granular Al, amorphous NbSi, discontinuous Au and amorphous InOx film. We can thus legitimately ask why they were not highlighted much earlier? Instead, in InOx films studies, it has been constantly repeated that no memory dip can be measured above few tens of K \cite{OvadyahuPRB1991, OvadyahuPRB1993, OvadyahuPRB2002, OvadyahuPRB2006b}. In the pioneer paper of 1991, it was written \cite{OvadyahuPRB1991}: ``At high $T$, regardless of $R_s$, the symmetrical component (the MD) does not appear. For example, a $1~G\Omega$ amorphous film measured at 77~K did not exhibit any symmetrical component, while at low $T$, such $R_s$ typically yield a 1 - 2~\% effect (1 - 4~K)''. This statement is in apparent contradiction with our findings. But according to the experimental results described previously, MDs are harder to identify at room $T$ than at 4~K for at least three reasons. First, they are overlapped by a linear field effect which amplitude can be larger than the MDs themselves, especially for low $R_s$ films. By comparison, the linear field effect remains relatively small at 4.2~K in granular Al films. Second, the MD amplitudes at room T in films usually measured in the liquid He range are small, typically less than 0.1~\% in granular Al films (see Fig.~\ref{Figure5}). These amplitudes are found to grow with $R_s$ and they reach values of the order of 1~\% in granular Al films only when $R_{s300K}$ lies above $1~G\Omega$. The measurement of large MDs at room $T$ thus requires the making of specific films, which due to the exponential-like divergence of the resistance at low $T$, can not be measured in the liquid He range. Third, the MDs at room $T$ are very wide compared to what they are at 4.2~K. In order to see them, it is therefore absolutely necessary to measure $G(V_g)$ curves over a large enough $V_g$ window. It was shown at 4.2~K that the MD dip width in a given sample is proportional to the thickness of the oxide layer used on the MOSFET device (it is the surface charge carrier density induced on the capacitor which matters \cite{OvadyahuPRB2002, OvadyahuPRB2008}). In our measurements, $V_g$ can be safely swept between -30~V and 30~V over 100~nm of $SiO_2$ and only the bottom of the dips are visible in all the films investigated (the FHWMs lie above 100~V in our granular Al films). In previous InOx studies, much smaller values were used: $\pm 100~V$ over $100~\mu m$ of glass \cite{OvadyahuPRB1991, OvadyahuPRB1993, OvadyahuEL1998, OvadyahuPRB2006a, OvadyahuPRB2006b} or $\pm 10~V$ over 500~nm of $SiO_2$ \cite{VakninPRL1998, OvadyahuPRB2014} respectively equivalent to $\pm 0.1~V$ and $\pm 2~V$ over 100~nm $SiO_2$. Such values do not allow the determination of MDs 100~V wide. In the granular Al film of Fig. 4 (right panel), a Lorentzian fit gives for the actual amplitude of the MD a relative value of 0.09~\%. The amplitude falls to 0.02~\% when measured between 0 and 30~V, to $10^{-4}$~\% between 0 and 2~V and much below between 0 and 0.1~V, i.e. definitely below the noise level of the experiment. This problem is evident at room $T$, but might also be important at lower $T$. In the $T$ study of InOx MD parameters published in Ref. \cite{OvadyahuEL1998}, the MD plotted on Fig. 1 is clearly larger than the explored $V_g$ range already at 4~K. An increase of the MD width with $T$, even if its actual amplitude is constant, will result in a decrease of its ``apparent'' amplitude. If no fit of the data with some known functions is performed (Lorentzian, …), it will not be possible to obtain a reliable temperature dependence for the MD amplitude. However, more recent $T$ studies stress that the MD of some amorphous InOx films displays a very sharp decrease around 4~K, even for a $T$ increase as small as 1~K \cite{OvadyahuPRB2008}. In granular Al films, a similar $T$ study was done over a larger $V_g$ range with the use of Lorentzian fits for the $G(V_g)$ curves \cite{GrenetEPJB2007}. A softer power law dependence was obtained for the MD amplitude decrease with $T$.

A further point we would like to emphasize is the fact that long relaxation times are present in all the films and systems measured, whatever $R_s$ and the MD amplitudes values. We found no evidence of some saturation to an equilibrium state, even after days of measurements, and thus no signs of a glass transition crossing between 4~K and room $T$. Instead, logarithmic time dependencies are observed, indicating that a wide distribution of relaxation times is still at work. The fact that the relaxation and MD relative amplitudes are smaller at room $T$ than at lower $T$ does not necessarily mean that the system is ``faster'' \cite{GrenetPRB2012}. This smallness can indicate that less slow modes are present in the system, but also that the conductance is less sensitive to the slow modes. In the absence of a more precise and quantitative model, it is not possible to conclude.

\section{Room temperature glassy effects and the electron glass hypothesis}

Lastly, we would like to discuss whether the room T glassy effects reported here can be explained by the electron glass hypothesis.

Most of the numerical simulations on Coulomb glass models agree on the existence of a slowdown of the dynamics as $T$ is lowered \cite{PerezPRB1999, MenasheEL2000, TsigankovPRB2003, GrempelEL2004, SomozaPRB2005, BergliPRB2011, PollakBookElectronGlass2012}. Such systems are characterized by a large number of metastable states, in which the transitions from one state to the other are believed to give rise, at low enough $T$, to relaxation times longer than any accessible experimental times \cite{PerezPRB1999, MenasheEL2000, TsigankovPRB2003}. The simultaneous hops of many electrons were found to play a crucial role in these very long relaxation times \cite{PerezPRB1999, TsigankovPRB2003, BergliPRB2011, AmirARCMP2011, OrtunoJPCS2012}. It was also suggested that the memory dip and its slow relaxation subsequent to $V_{geq}$ changes are due to the slow rearrangement of many-electron clusters, which modify the occupation probability of the electronic sites participating to the transport \cite{KozubPRB2008, AmirARCMP2011}. As to the existence of a true thermodynamic glass transition, the issue remains theoretically controversial. Mean field theories have predicted its existence in 3D and 2D systems \cite{MullerPRL2004, PankovPRL2005, MullerPRB2007}. $T_g$ estimates were found to be strongly dependent on the system parameters, and possibly larger than the Coulomb gap energy in 2D systems having strong disorder and large charge carrier densities \cite{LebanonPRB2005}. Numerical investigations found either no trace of such a transition \cite{GoethePRL2009}, or $T_g$ values much smaller than the mean field predictions \cite{BarzegarPRB2019}.

A feature common to all these models is that electronic glassy effects manifest at temperatures smaller than the Coulomb gap energy $U_{CG}$, which is the typical energy scale below which electronic correlations become important. In a disordered system, $U_{CG}$ is given in 3D by $\alpha e^3(N(E_F))^{1/2}(4\pi\epsilon)^{-3/2}$, where $N(E_F)$ is the density of states at the Fermi level, $\epsilon$ the effective dielectric constant and $\alpha$ a numerical constant close to unity \cite{EfrosJPC1975}. In our InOx film, the MD width at 17~K (the lowest temperature at which the $G(V_g)$ curves were still measurable, $R = 17~G\Omega$) is equal to $\simeq 4~V$, which gives by extrapolation a MD width of $\simeq 1~V$ at 4~K. By using the correlation between the charge carrier concentration $n$ and the 4~K MD width reported in InOx films \cite{VakninPRL1998}, we find that $n \approx 10^{20}~e/cm^3$. Taking the free electron formula for $N(E_F)$ and a relative dielectric constant of 10 \cite{OvadyahuPRB1991}, we finally obtain $U_{CG} \approx 0.1~eV$. In NbSi films, the charge carrier density is not known precisely but it may be larger than $10^{22}e/cm^3$ \cite{DelahayeSciPost2020}, which will correspond to a Coulomb gap charging energy above 0.2~eV (again assuming a relative dielectric constant of 10). In granular Al films, $U_{CG}$ should be of the order of the charging energy of the Al grains. For a typical grain diameter of 5~nm and a tunnel barrier 1~nm-thick, we get a charging energy of the order of 100~K \cite{GrenetEPJB2007}. We should bear in mind that all these estimates have a high level of uncertainty: the relative dielectric constant is not precisely known in amorphous InOx and NbSi films, there is a distribution of grain size and tunnelling barriers in granular Al, and these parameters may change strongly at the approach of the metal-insulator transition. Energy distributions extending above 300~K thus seem plausible.

One may also consider activation energies extracted from transport and slow relaxation experiments. In insulating granular Al films the resistance diverges almost exponentially as $T$ is lowered. Using it, we get a wide range of activation energies depending on the insulating character of the samples : from 1000~K when $R_{s300K}\approx 100~G\Omega$ to 100~K when $R_{s300K}\approx 100~k\Omega$. In granular Al films the MD dynamics was shown to depend on $T$ in a complex manner, showing effective activation energies which depend on the temperatures involved in the measurements, and can be as high as 2000~K when the dynamics is measured around 50~K \cite{GrenetVillard2018}. A similar behaviour was also recently observed in InOx \cite{GrenetVillard2018}.

However the picture is not that simple.
First, the case of discontinuous Au is peculiar. This material presents a wide distribution of Au island size and shape (see Fig.~\ref{Figure0}) but many of them stretch on more than 100~nnm, which would imply charging energies of less than $\simeq 10~K$. Activation energies estimates from resistance measurements are also smaller than in granular Al : for example less than 10~K when $R_{s300K} = 100~M\Omega$.
Second, according to a percolation approach of the memory effects, the MD width is expected to be proportional to $k_BT$ when the latter is larger than the coulomb gap \cite{LebanonPRB2005}. Since this temperature dependence is approximately observed down to 4~K in granular Al, InOx and NbSi, one may deduce that $U_{CG}$ is less than 1~meV !

Regardless of any specific model consideration, relaxation times as long as days ($\approx 10^5~s$) at room $T$ involve processes with activation energies as high as 0.5–1~eV (attempt rates between $10^{12}$ to $10^6~s^{-1}$). In an intrinsic picture where these slow relaxations result from the rearrangement of electron clusters \cite{KozubPRB2008, AmirARCMP2011}, it implies either individual electron hops with activation energies of 0.5-1~eV, at the upper limit of our Coulomb gap energy estimates, or simultaneous hops of $N$ electrons, each of it having a smaller average energy \cite{KozubPRB2008}. Whether such collective events are indeed present in significant number at room $T$ in our disordered films, especially not too far from the metal-insulator transition and in discontinuous Au films, deserves further theoretical and numerical studies.

\section{Conclusion}

We have identified for the first time electrical glassy features at room $T$ in granular Al and InOx films. MDs are present in all the films we have been measuring. Their $V_g$-widths are about 100 times larger than what is usually found at 4~K, while their time evolutions under $V_{geq}$ changes obey similar logarithmic dependences. The existence of room $T$ glassy effects, so far only mentioned in discontinuous Au and NbSi films, seems to be the rule rather than the exception among electron glass systems.

Our study clarifies what experimental conditions are required in order to measure room $T$ glassy effects. In particular, the large $V_g$ widths of the MDs at room $T$ require the use of a large enough $V_g$ range in field effect measurement (or of a thin enough gate insulating material), a condition which was not fulfilled in previous attempts on InOx films. Some specific difficulties were also met: in discontinuous Au films, the MDs become measurable only a few hours after the making of the films, and in amorphous InOx films, the usual conductance relaxation is mixed with a relaxation mechanism of different nature. More extensive investigations are needed in order to clarify the physical origin of these two features.

The existence and the ability to measure electrical glassy effects up to room $T$ open promising perspectives. First, it allows the use of new experimental techniques, such as local electrical probes, more difficult to implement in the liquid He range. Second, the knowledge of the MD evolution between 4~K and 300~K must help to build quantitative models of the electrical glassy effects and to precise the discrepancies/similarities between the different systems. In particular, a surprising $T$ dependence was recently demonstrated for the glassy dynamics of granular Al and InOx films below 30~K \cite{GrenetJPCM2017, GrenetVillard2018}, with an increase of the apparent activation energy with $T$. It will be interesting to determine how the glassy dynamics evolves at higher $T$.

\section*{Acknowledgements}
We acknowledge Claire Marrache-Kikuchi, Laurent Berg\'e and  Louis Dumoulin for providing NbSi samples and for their comments on the article.
We also gratefully acknowledge discussions with Markus M\"uller.

\end{document}